\documentclass[]{jfm}

\usepackage{graphicx}
\usepackage{newtxtext}
\usepackage{newtxmath}
\usepackage{natbib}
\usepackage{hyperref}
\hypersetup{
    colorlinks = true,
    urlcolor   = blue,
    citecolor  = black,
}

\newcommand{\RomanNumeralCaps}[1]
\linenumbers

\usepackage{xcolor}

\newcommand{\reva}[1]{\textcolor{black}{#1}}
\newcommand{\revb}[1]{\textcolor{black}{#1}}

\title{$2.5$D turbulence in shear-thinning jets}

\author{Christian Amor\aff{1},
        Giovanni Soligo\aff{1},
        Andrea Mazzino\aff{2,3}
        \and Marco E. Rosti\aff{1}
        \corresp{\email{marco.rosti@oist.jp}}}

\affiliation{\aff{1}Complex Fluids and Flows Unit, Okinawa Institute of Science and Technology Graduate University (OIST), 1919-1 Tancha, Onna-son, Okinawa 904-0495, Japan
\aff{2}Department of Civil, Chemical and Environmental Engineering, Via Montallegro 1, Genova, 16145, Italy
\aff{3}INFN, Instituto Nazionale di Fisica Nucleare, Sezione di Genova, Via Dodecaneso 33, Genova, 16146, Italy}

\begin{document}
\maketitle

\begin{abstract}
The dimensional transition in turbulent jets of a shear-thinning fluid is studied via direct numerical simulations. Our findings reveal that under vertical confinement, the flow exhibits a unique mixed-dimensional (or $2.5$D) state, where large-scale two-dimensional and small-scale three-dimensional structures coexist. This transition from three-dimensional turbulence near the inlet to two-dimensional dynamics downstream is dictated by the level of confinement: weak confinement guarantees turbulence to remain three-dimensional, whereas strong confinement forces the transition to two-dimensions; the mixed-dimensional state is observed for moderate confinement and it emerges as soon as flow scales are larger than the vertical length. In this scenario, we observed that the mixed-dimensional state is an overall more energetic state and it shows a multi-cascade process, where the direct cascade of energy at small scales and the direct cascade of enstrophy at large scales coexist. The results provide insights into the complex dynamics of confined turbulent flows, relevant in both natural and industrial settings.
\end{abstract}

\begin{keywords}
Jets, shear-layer turbulence, turbulence simulation
\end{keywords}

\section{Introduction}\label{sec:intro}

The flow of a low-viscosity fluid at high-speed is chaotic in nature. The energy injected to sustain this state is transferred from large to small eddies, down to a particular scale from which it is dissipated by the viscosity of the fluid \citep{kolmogorov1941local}. Conventional turbulence in three-dimensions fulfills this description, whereas new phenomena appear in two-dimensions. Energy transfer in two-dimensional turbulence is dictated by a double cascade scenario: an inverse cascade of kinetic energy to large scales and a direct cascade of enstrophy (squared vorticity) to small scales \citep{kraichnan1967inertial,batchelor1969computation,boffetta2007energy}. Certainly, no physical system is two-dimensional in reality, though two-dimensional turbulence becomes relevant if one spatial direction is greatly constrained, e.g., by geometry \citep{boffetta2012bolgiano,boffetta2012two}. For instance, the large-scale motions in atmospheric flows comply with two-dimensional turbulence \citep{charney1971geostrophic,nastrom1984kinetic,lindborg1999can}. In this case, the flow domain is subdued to a large aspect ratio: the horizontal lengths are much larger than the height of the atmospheric layer. 

The confinement in thin layers can induce a rich phenomenology in turbulent flows that, if forced at intermediate scales, produces a split energy cascade \citep{smith1996crossover,celani2010turbulence,alexakis2018cascades}. Under this circumstance, a portion of the energy flows to the large scales in a two-dimensional-fashion. Conversely, the remaining part cascades toward the small viscous scales. Interestingly, a direct cascade of enstrophy can develop simultaneously at scales smaller than the forcing but larger than the thickness of the layer, and three-dimensionality becomes relevant only at much smaller scales \citep{musacchio2017split}. Nevertheless, the presence of physical confinement, e.g., using solid boundaries, is not compulsory to observe the split energy cascade. In fact, numerical simulations in a fully-periodic box with one dimension much smaller than the others have shown this phenomena \citep{smith1996crossover,celani2010turbulence}. Despite this, its occurrence changes with the boundary conditions. For example, the development of the shear layer in wall-bounded flows restricts the development of two-dimensional dynamics \citep{xia2011upscale,byrne2011robust,boffetta2023transient}. 

Here, we consider a planar jet, i.e., the flow is injected through a plane slit of half-width $h$ in a computational box periodic in the vertical direction $z$. We adopt a shear-thinning fluid in which the viscosity decreases non-linearly for increasing values of the shear rate.
A similar flow configuration, although at a much higher characteristic Reynolds number, can be found at the outflow of a river into the sea. Differences in salinity, temperature and density between the freshwater stream and the salt water can impede mixing, thus leading to the formation of a stratified flow with a (thin) layer of freshwater flowing over salt water. The presence of suspended bacteria and microalgae in the freshwater stream grants non-Newtonian features to the fluid, such as shear-dependent viscosity \citep{al2002rheological,zhang2013influence}.
Our simulations are performed at a much lower value of the Reynolds number, hence there is no direct application of our findings to the flowing regimes found at the outflow of a river; the low value of the Reynolds number allows instead for direct comparisons with experiments. The present configuration can be easily achieved in laboratory experiments; the shear rheology of the fluid used in our simulations corresponds to that of a 100:60 mM CPyCl:NaSal wormlike micelle solution \citep{haward2021stagnation}. This work thus constitutes a preliminary step in understanding the flow dynamics of geophysical flows, such as the outflow of a river rich in bacteria or microalgae, which are instead characterized by a much larger Reynolds number. The selected parameters make our numerical setup easily testable by experiments (recent experimental work by \citet{yamani2023spatiotemporal} addressed the flow of a viscoelastic planar jet at low Reynolds), whereas the effect of physical confinement is attenuated by the shear-thinning characteristic of the fluid, which reduces the extent of the shear layer (viscosity decreases at the wall boundaries in the experimental setup).

Within this framework, we show that when the thickness of the domain is large, the flow is completely three-dimensional, while when it is small, it is fully two-dimensional. Interestingly, for intermediate cases, the flow spatially transitions from $3$D close to the inlet to $2$D further downstream, with the two regimes being connected by a region of mixed-dimensional turbulent dynamics where the constraint modulates the largest scales towards two-dimensions and the smaller ones remain three-dimensional. We indeed observe, at intermediate levels of vertical constraint, a multi-cascade process, where both a direct cascade of energy at small scales and a direct cascade of enstrophy at large scales coexist.

\section{Numerical method}

The motion of the incompressible, shear-thinning fluid is governed by the mass, Eq.~\ref{eq:cont}, and momentum, Eq.~\ref{eq:ns}, conservation equations. 
\begin{equation}
\nabla \cdot \boldsymbol{u} = 0
\label{eq:cont}
\end{equation}
\begin{equation}
\rho \left(\frac{\partial \mathbf{u}}{\partial t} + \mathbf{u}\cdot\nabla{\mathbf{u}}\right) = -\nabla p + \nabla \cdot \left[ \mu(\dot\gamma) \left( \nabla\mathbf{u}+\nabla\mathbf{u}^T \right) \right] 
\label{eq:ns}
\end{equation}
In the above equations, $\mathbf{u}$ is the local flow velocity, $\rho$ the density, $p$ the pressure and $\mu$ the local viscosity.
We adopt an inelastic, shear-thinning fluid, whose behaviour is defined via the Carreau fluid model \citep{bird1974co}. The local viscosity $\mu$ depends on the local shear rate $\dot\gamma$ as:
\begin{equation}
	\mu(\dot\gamma) = \mu_{\infty} + \left(\mu_0 - \mu_{\infty}\right)\left[1 + \left(\lambda \dot{\gamma}\right)^{2}\right]^{\frac{n-1}{2}},
	\label{eq:a1}
\end{equation}
where $\lambda$ is the fluid consistency index, and $\mu_0$ and $\mu_{\infty}$ are the zero-shear viscosity and the viscosity for $\dot{\gamma} \rightarrow \infty$, respectively. We set the power-law index $n = 0.2$, thereby obtaining a strong shear-thinning effect. The local shear rate is defined as $\dot\gamma=\sqrt{2~\mathsfbi{S}:\mathsfbi{S}}$, where $\mathsfbi{S}=(\nabla\mathbf{u}+\nabla\mathbf{u}^T)/2$ is the shear rate tensor.

The Navier-Stokes equations are discretized on a uniform, staggered, Cartesian grid; the fluid velocities are located at the cell faces, whereas pressure and viscosity are defined at the cell centers. The fluid viscosity is updated at every time-step following Eq. \ref{eq:a1}. The spatial derivatives are approximated using second-order finite differences in all directions. The system is advanced in time through a second-order Adams-Bashforth scheme coupled with a fractional step method \citep{kim1985application}. The divergence-free velocity field is enforced by solving the Poisson equation for the pressure using an efficient solver based on the fast Fourier transform. We resort to the domain decomposition library 2decomp (\url{http://www.2decomp.org}) and the MPI protocol to parallelize the solver. The numerical solver is implemented in the in-house solver \textit{Fujin} (\url{https://groups.oist.jp/cffu/code}). 
 
\section{Problem setup}\label{sec:setup}

We have addressed this study by means of three-dimensional direct numerical simulations; the computational box has sizes $L_x=160h$ in the stream-wise direction, $L_y=240h$ in the jet-normal direction and $0.83h \le L_z \le 13.33h$ in the span-wise direction. 
The vertical length $L_z$ is varied among simulations: we consider five distinct simulations with $L_z = 0.83h, 1.67h, 3.33h, 6.67h, 13.33h$, respectively. The thinnest domain ($L_z = 0.83h$) introduces a strong vertical constraint in order to allow the development of a two-dimensional flow, which is progressively relaxed as $L_z$ is increased while still maintaining a thin computational domain, so $L_z \ll L_x \sim L_y$.
The planar jet is generated by fluid injected with a uniform velocity $U$ through a planar slit of width $2h$ spanning the entire height of the domain ($z$ direction).
At the inlet boundary we impose no-slip and no-penetration boundary conditions, exception made for the inlet portion, where we impose a plug-flow velocity profile. At the outlet boundary ($x = L_x$) we use a non-reflective boundary condition \citep{orlanski1976simple}. At the side boundaries ($y = 0$ and $y = L_y$) we impose free-slip and no penetration boundary conditions. Lastly, at the top and bottom boundaries ($z = 0$ and $z = L_z$) we impose periodic boundary conditions.

We select a low value of the inlet Reynolds number (ratio of inertial to viscous effects), \reva{$Re = \rho h U / \mu_{0} = 20$}. It should be noted that Newtonian planar jets are laminar at this value of $Re$ \citep{deo2008influence,sureshkumar1995effect,sato1964experimental,soligo2023non}, 
thus any turbulent motion is caused exclusively by the shear-thinning in the flow. Turbulence however is still Newtonian, as it originates by the prevalence of inertial over viscous terms. The non-Newtonian character of the flow promotes indeed the onset of the instability, so the transition to turbulence is at markedly lower $Re$ compared to Newtonian planar jets \citep{ray2015absolute,yamani2023spatiotemporal,soligo2023non}. The non-Newtonian contribution is described using the Carreau number, defined as $Cu = h \lambda / U = 100$. 
The ratio between the zero-shear viscosity and the infinite-shear viscosity is set to $\mu_0 / \mu_{\infty} = 50$.

To verify the independence of our results on the specific parameters selected, we perform two additional simulations at a set $L_z=3.33h$ and we double either the Reynolds ($Re = 40$) or the Carreau ($Cu = 200$) number. The reference case ($L_z=3.33h$, $Cu=100$ and $Re=20$) exhibits a mixed-dimensional turbulent regime, where features from two- and three-dimensional turbulence are found simultaneously in the flow (see Fig. \ref{fig:ene}).
In the cases at higher $Re$ or $Cu$, we expect turbulence to be enhanced. The Reynolds number is increased by reducing the zero-shear viscosity (and consequently the infinite-shear viscosity, defined as $ \mu_{\infty} = \mu_0 /50$), whereas the Carreau number is increased by changing the fluid consistency index $\lambda$, keeping the zero-shear and infinite shear viscosities unchanged. This way, the transition towards the infinite-shear viscosity occurs at a smaller shear rate compared to the case at $Cu=100$. 


We adopt a uniform grid spacing in all spatial directions for all cases; in $x$ and $y$ the domain is discretized using $N_x \times N_y = 1536 \times 2304$ grid points, while the number of points in the $z$ direction depends on the height of the domain, namely $N_z = 8, 16, 32, 64, 128$ for increasing heights. 
We ensured that the grid resolution is adequate by computing the ratio between the grid spacing $\Delta$ and the mean Kolmogorov length scale $\eta$:
\begin{equation}
    \eta = \left( \frac{\langle \nu \rangle^3}{\langle \varepsilon \rangle} \right)^{1/4},
    \label{eq:kolm}
\end{equation}
where $\nu$ is the local kinematic viscosity and $\varepsilon$ is the viscous dissipation; angle brackets indicate averaging in time and in the vertical direction $z$.
In the cases characterized by the highest turbulence intensity, $Re=40$ and $Cu=200$, the Kolmogorov length scale is always $\eta \gtrsim 0.5 \Delta$ (where $\Delta$ is the grid spacing, uniform in the three directions). The smallest values of the Kolmogorov length scale are encountered at the jet centerline within the region $15h < x < 30h$; beyond $x = 40h$ the Kolmogorov length scale is always larger than the grid spacing, thus ensuring that the grid resolution chosen is adequate for all cases.

\section{Results} 
\subsection{Effect of vertical confinement}
\label{sec:resLz}
\begin{figure}
    \centering	\includegraphics[width=\textwidth,height=\textheight,keepaspectratio]{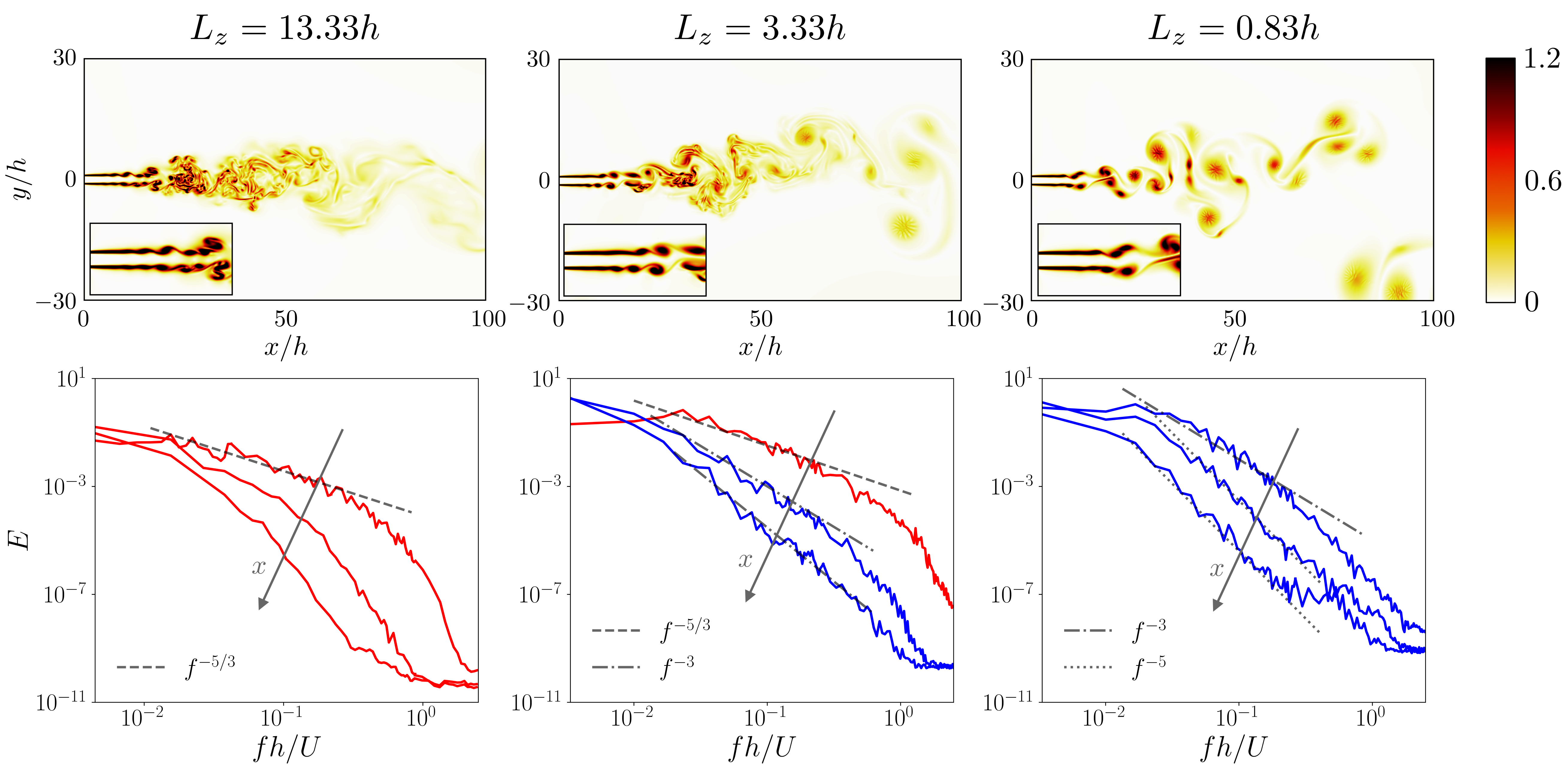}
    \caption{\label{fig:ene} 
    Effect of the constrained dimension $L_z$ in the turbulent planar jets. Top panel: magnitude of the instantaneous vorticity field $\left|\left| \boldsymbol{\omega} \right|\right| U^{-1} h$ 
    in the $z = L_z/2$ plane for the three-dimensional, mixed-dimensional and two-dimensional planar jets, from left to right. The inset shows a zoomed view of the region near the inlet at $x = \left[ 0, 20h \right]$, $y = \left[ -5h, 5h \right]$. Bottom panel: power spectra of the turbulent kinetic energy computed at the jet centerline at distances $x = 40h, 80h, 120h$ from the inlet. The spectra follows the typical $f^{-5/3}$ scaling if turbulence is primarily three-dimensional (red). In two-dimensional turbulence (blue), instead, the flow exhibits a $f^{-3}$ decay that becomes steeper as $x$ increases, tending towards $f^{-5}$ in the most constrained planar jet.}	
\end{figure}
Figure \ref{fig:ene} shows the impact of $L_z$ on the turbulent planar jets, in which, as anticipated, the constraint dictates the turbulent regime in the flow. Consequently, the morphology of the vorticity structures changes significantly with $L_z$ (see top panel in Fig. \ref{fig:ene}). We first observe that turbulence is three-dimensional if the flow is not constrained, i.e., for sufficiently large values of $L_z$. A complete different flow structure is instead observed at low values of $L_z$, in which large vortices form dipoles (pairs of counter-rotating vortices) that are advected downstream, and no small scale flow structures are observed, thus indicating that turbulence is mainly two-dimensional. The flow does not transition in bulk from $3$D to $2$D when changing $L_z$, with the planar jet experiencing an intermediate state where both three-dimensional and two-dimensional structures are present in the flow at the same time. This transitional regime, hereafter termed \textit{mixed-dimensional} (or $2.5$D), is characterized by the simultaneous coexistence in the flow of large-scale two-dimensional and small-scale three-dimensional structures.

Next, we inspect the energy spectra in the bottom panel of Fig. \ref{fig:ene}. Note that we compute the velocity spectra in time rather than in space by recording velocity data over time from a probe placed on the centerline of the jet, similarly to what done in experiments. Computing the power spectra in time rather than in the vertical direction allows us to have a wider energy spectrum that is not limited by the height of the domain. The equivalence of time and space spectra has been demonstrated in the past \citep{namer1988velocity,soligo2023non}. As clearly shown in the figure, $L_z$ influences the energy cascade, which depicts a different behavior depending on the regime of turbulence. First, the least constrained jet shows the conventional $f^{-5/3}$ scaling for three-dimensional turbulence \citep{kolmogorov1941local}. Consistently, we observe the energy cascade typical of three-dimensional turbulence: the jet instability gives energy to the flow and it generates large structures that break down in progressively smaller and smaller eddies. As the vortices move downstream, the characteristic shear rate reduces -- hence viscosity increases -- and energy is dissipated. We expect to recover the power-law scaling for the three-dimensional turbulence energy cascade as the turbulent motions are Newtonian: they are generated by the competition of inertial and viscous terms \citep{soligo2023non}. Eventually, dissipation becomes relevant at every scale and the cascade is impeded: the scaling $f^{-5/3}$ is not present at $x = 120h$. On the other hand, the most constrained case exhibits two-dimensional flow and a scaling of $f^{-3}$. Here, the three-dimensional cascade is clearly disrupted and two-dimensional phenomena become dominant \citep{kraichnan1967inertial,batchelor1969computation}. The change in the flowing regime observed here is due only to the vertical confinement: we adopt a non-Newtonian fluid model which is characterized by the presence of shear-thinning alone (there are no viscoelastic effects). The spectrum becomes steeper as $x$ increases and it eventually seems to saturate at $f^{-5}$. 
The steepening of the energy spectrum agrees with the appearance of dispersed large-size coherent vortices in the flow 
\citep{basdevant1981study,mcwilliams1984emergence,benzi1986intermittency,legras1988high}. Note that the change in the vertical constraint also alters the instability in the region close to the inlet: for strong vertical constraint (small $L_z$) we observe a flapping motion of the shear layers, whereas puffing motion dominates when the constraint is relaxed (large $L_z$). The flapping dynamic is associated with the antisymmetric, sinuous mode, that destabilizes the flow more substantially, thus injecting more energy \citep{mattingly1971disturbance}, as also observed by the energy spectra close to the inlet, which are shifted upwards in 2D compared to the 3D case. At last, the mixed-dimensional planar jet exhibits features from both three- and two- dimensional planar jets: the scaling $f^{-5/3}$ is found close to the inlet and it changes towards $f^{-3}$ downstream. The spectrum becomes slightly steeper further downstream, consistent with the appearance of the coherent vortices. Similarly to the two-dimensional jet, the flow is more energetic close to the inlet. Very close to the inlet, the vorticity fluctuations look closely related to those in the three-dimensional case, implying the existence of puffing events connected to the varicose mode \citep{sato1960stability,mattingly1971disturbance}. However, flapping motions become soon dominant and the macroscopic vorticity structures resemble those from the two-dimensional planar jet, thus indicating the presence of the more energetic sinuous mode. 

\begin{figure}
    \centering
	\includegraphics[scale=0.5,keepaspectratio]{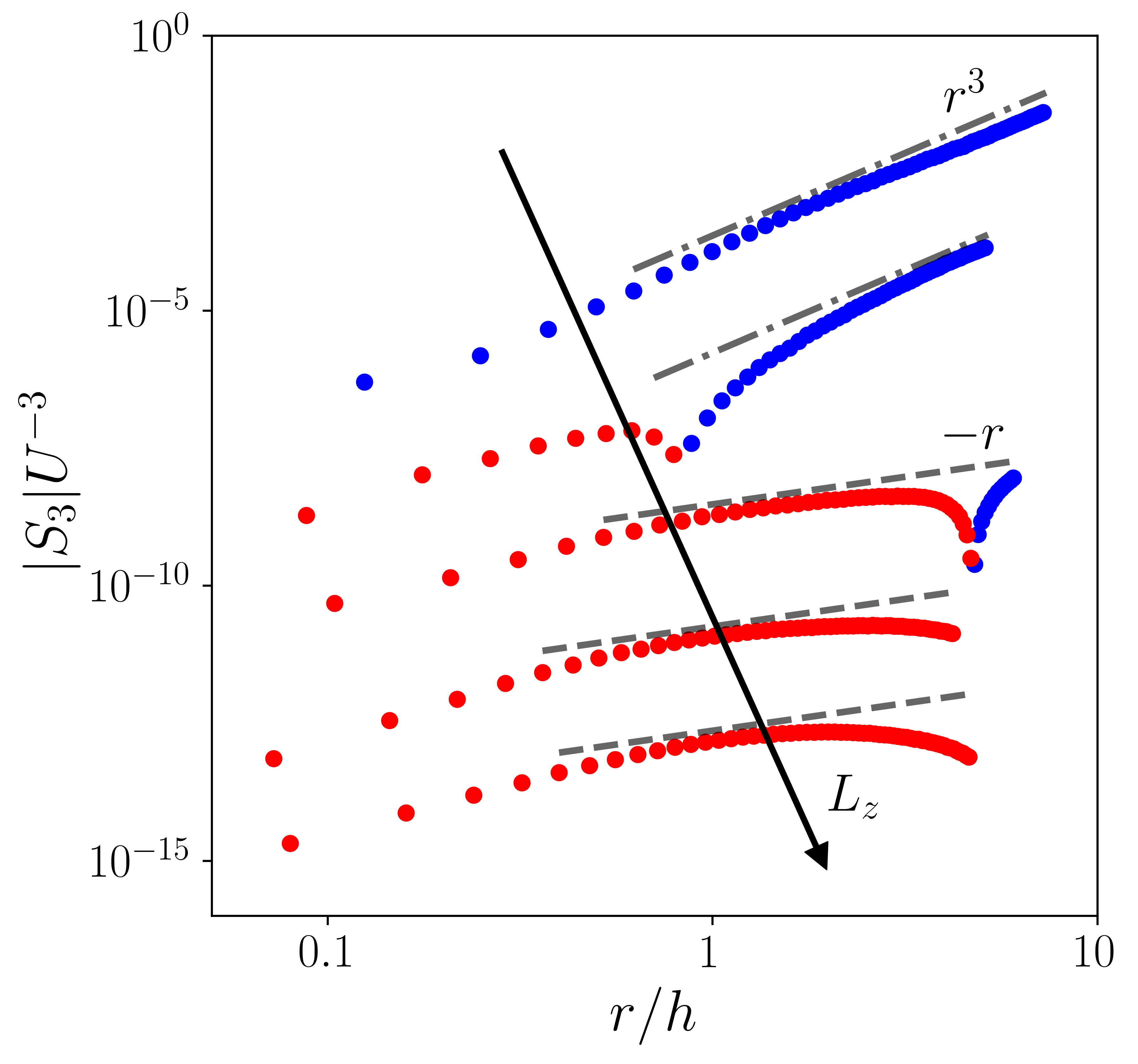}%
    \caption{\label{fig:strfunall} 
    Third-order structure functions $S_3$ of the longitudinal velocity fluctuations for different vertical constraints $L_z$. We compute $S_3$ at $x = 40h$ at the flow centerline for each planar jet: $L_z = 0.83h, 1.67h, 3.33h, 6.67h, 13.33h$. We show the absolute value $\left| S_3 \right|$ for clarity, and we indicate with colors whether $S_3$ is positive (blue) or negative (red), so that the corresponding turbulent scales are either two- or three- dimensional. Consequently, the scaling for $\left| S_3 \right|$ changes as $L_z$ is shifted towards larger values. We report also the two scalings: $S_3 \sim r^3$ if turbulence is two-dimensional, corresponding to a direct cascade of enstrophy, and $S_3 \sim -r$ if it is three-dimensional, thus denoting a direct cascade of energy. The plots are shifted vertically for better readability.}	
\end{figure}
To better characterise the different nature of the turbulent fluctuations at each scale, we now calculate the longitudinal velocity differences $\Delta u(r) = \left( \boldsymbol{u} \left( \boldsymbol{x} + \boldsymbol{r} \right) - \boldsymbol{u} \left( \boldsymbol{x} \right) \right) \cdot \boldsymbol{r} / \left| \boldsymbol{r} \right|$. Concretely, we introduce the third-order structure function $S_3(r) = \langle \left( \Delta u \right)^3 \rangle$, where the angle brackets indicate averaging in time and in space, shown in Fig. \ref{fig:strfunall} for the different cases analysed. Appendix~\ref{app:strucfun} reports in detail how the computation of the structure function was performed.
A remarkable property of $S_3$ is that it denotes whether the flow scales are two- or three- dimensional depending on its sign \citep{lindborg1999can,kolmogorov1991dissipation}, and it can help understanding the direction of the energy and enstrophy fluxes \citep{cho2001horizontal,bernard1999three,cerbus2017third}. In the particular case of two-dimensional turbulence, $S_3$ is positive \citep{lindborg1999can} whereas it is negative if turbulence is three-dimensional \citep{kolmogorov1991dissipation}.
Indeed, we observe that $S_3$ is positive (two-dimensional flow) for the most constrained case ($L_z = 0.83h$) whereas it is negative (three-dimensional flow) for the least constrained case ($L_z = 6.67h, 13.33h$). Furthermore, the scaling of $S_3$ outlines the preferred cascade process, thus indicating the direct cascade of enstrophy in the two-dimensional case ($S_3 \sim r^3$) and the direct cascade of energy in the three-dimensional counterpart ($S_3 \sim -r$). 
While the scaling $S_3 \sim r^3$ is visible for two-dimensional turbulence, we do not observe a clear $S_3 \sim -r$ scaling for all three-dimensional turbulence cases.
We do not find evidence of an inverse mechanism of energy transfer toward the large scales in the two-dimensional case. This is not surprising if we consider that energy is injected in the flow through the planar slit, thus not forcing the flow at any intermediate scale \citep{boffetta2007energy,boffetta2012two}. More interestingly, turbulence is characterized by a mixed-dimensional regime for the right choice of $L_z$. Two- and three- dimensional scales are present simultaneously in the flow for $L_z = 1.67h, 3.33h$. In these cases, large scales are two-dimensional while the small ones are three-dimensional, with the transition between regimes happening at some intermediate scale. This transition is strongly dependent on the height of the domain and it can be delayed further in $x$ for increasing values of $L_z$. 

\begin{figure}
    \centering	\includegraphics[scale=0.25,keepaspectratio]{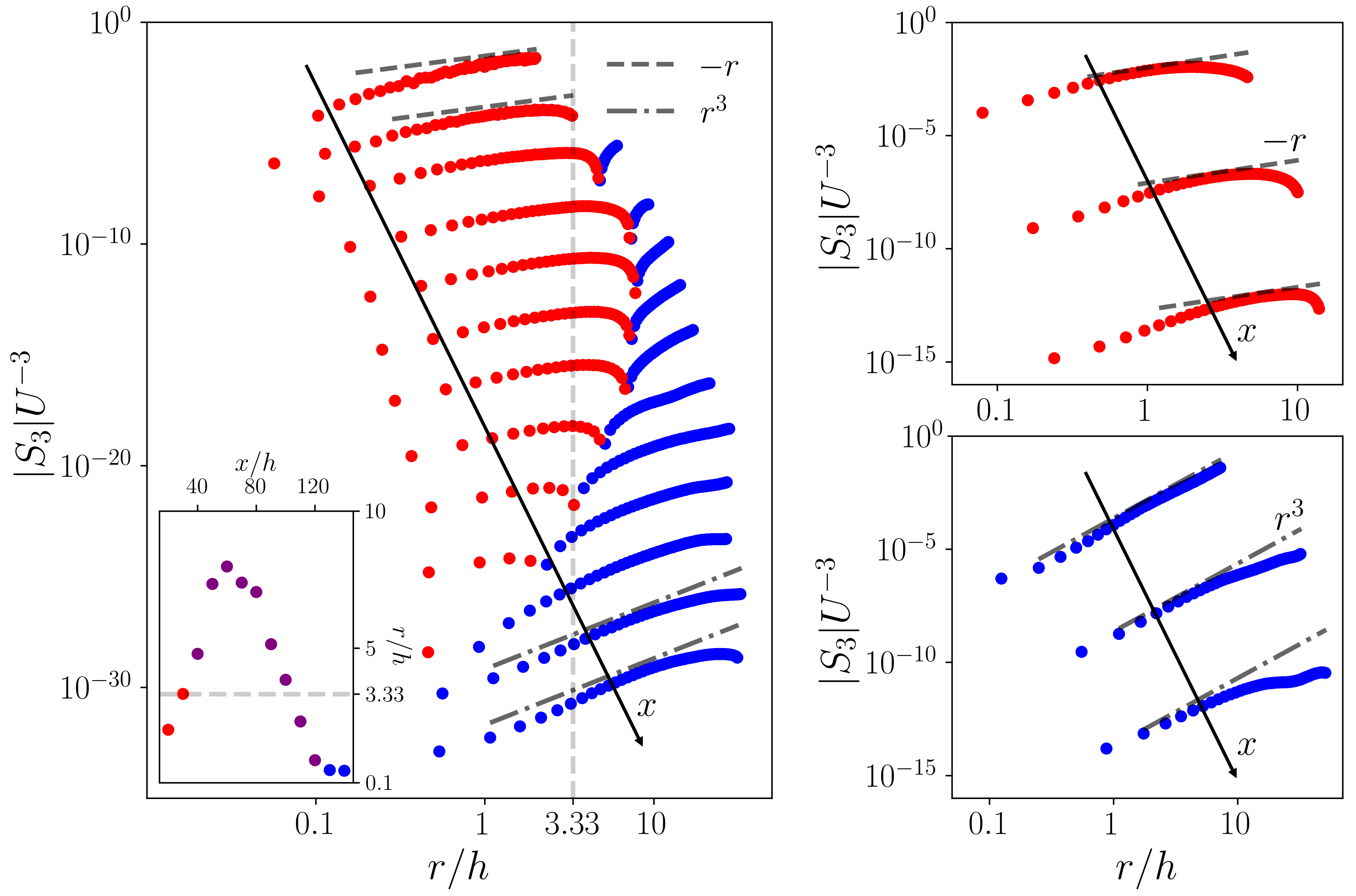}%
    \put(-360,235){\large \textit{a})}
    \put(-145,235){\large \textit{b})}
    \put(-145,123){\large \textit{c})}
    \caption{Third-order structure functions $S_3$ of the longitudinal velocity fluctuations at different distances from the inlet. Panel \textit{a} shows $S_3$ computed at distances $x$ uniformly separated along the jet centerline ($x = 20h, 30h, \ldots, 140h$) for the $2.5$D turbulent planar jet with $L_z = 3.33h$. We indicate whether the turbulent scales are two- or three- dimensional with colors, similarly to Fig. \ref{fig:strfunall}. The plots are shifted vertically for better readability. The inset shows the scale $r$ at which $S_3$ changes its sign. Red and blue markers are used for fully $3$D and $2$D structure functions respectively, while purple markers for those showing both regimes. Note that, if $S_3$ does not change sign, we mark either the largest three-dimensional scale or the smallest two-dimensional one. Panel \textit{b} depicts $S_3$ for the least constrained ($L_z = 13.33h$) or $3$D jet, and panel \textit{c} for the most constrained ($L_z = 0.83h$) or $2$D jet. $S_3$ are plotted in a similar fashion than \textit{a}, where we computed $S_3$ at three distances $x = 40h, 80h, 120h$ at the jet centerline. In these cases, $S_3$ has the same sign throughout $x$, thus displaying a single scaling at all distances dependent on the dimensionality of the flow.}	
    \label{fig:strfun32} 
\end{figure}
To further investigate the transition between regimes, we choose a mixed-dimensional planar jet, in particular $L_z = 3.33h$, and we calculate repeatedly $S_3$ at several distances from the inlet. Results are summarized in Fig. \ref{fig:strfun32}\textit{a}. Close to the inlet the flow is three-dimensional: the effect of the vertical constraint does not hinder the development of the three-dimensional regime. The largest flow scale is smaller than the geometrical constraint, thus not compromising the energy cascade of the three-dimensional flow. We report the presence of a single scaling, $S_3 \sim -r$, typical of the direct energy cascade. At intermediate distances, $x = \left[ 40h, 100h \right]$, a mixed-dimensional regime appears: both two- and three-dimensional regimes coexist, with three-dimensional turbulence characterizing the small scales and two-dimensional turbulence characterizing instead the large scales. Here, the vertical confinement hinders the development of three-dimensional turbulence at the largest scales, while it has no effect on the three dimensional turbulence at the smallest scales. In this case, we have the simultaneous presence of a direct energy cascade ($S_3 \sim -r$) at small scales and a direct enstrophy cascade ($S_3 \sim r^3$) at large scales. We report in the inset of Fig. \ref{fig:strfun32}\textit{a} the scale at which the structure function changes sign, i.e., the scale where the flow transitions from $3$D to $2$D, so the largest $3$D scale in the flow. We observe that, as soon as the jet thickness\footnote{The jet thickness is defined as the distance from the centerline at which the streamwise velocity equals half of the centerline velocity.}, which corresponds to the largest $r$, reaches the size of the vertical confinement, the largest scales become two-dimensional. However, some scales between the largest ones and the vertical confinement $L_z$ remain three-dimensional, causing the largest $3$D structures to be anisotropic. Moving downstream, the anisotropy of the largest $3$D structures grows till a value of around $3L_z$ at $x\approx 60h$, after which the flow becomes more and more two-dimensional. Eventually, we observe that the flow becomes completely two-dimensional at the farthest distances from the inlet, $x \geq 120h$, where the characteristic flow scales are the largest and the vertical confinement impedes the development of three-dimensional flow at all scales. We indeed observe a scaling of $S_3$ compatible with the direct enstrophy cascade ($S_3 \sim r^3$). The range of scales observed at each distance from the inlet depends on two factors: the characteristic length scale of the jet (the jet thickness) and the local viscosity. Both of these quantities increase with increasing distance from the inlet of the jet; the jet thickness determines the largest scale in the flow, while the local viscosity is among the factors determining the eventual development of three-dimensional turbulent motions.

Lastly, we perform the same analysis on the least ($L_z = 13.33h$) and the most ($L_z = 0.83$) constrained jets at several locations on the jet centerline ($x = 40h, 80h, 120h$) (see Fig. \ref{fig:strfun32}\textit{b}, \textit{c}). In these cases, $S_3$ has the same sign throughout $x$, indicating that turbulence is either three- ($S_3 < 0$) or two- ($S_3 > 0$) dimensional in the planar jet. 
For the three-dimensional planar jet, the structure function approaches the scaling for the direct cascade of energy, $S_3 \sim -r$, at intermediate values of the separation distance $r$.
On the other hand, the scaling for $S_3$ in the two-dimensional planar jet shifts towards $S_3 \sim r^3$, indicating instead the presence of the direct cascade of enstrophy. More interestingly, the scaling holds for small separation distances further from the inlet. This is in good agreement with the concept of the enstrophy cascade as a space-filling phenomenon, thus being present at very small scales \citep{benzi1986intermittency}. Additionally, $S_3$ follows an anomalous behavior at large and intermediate $r$. As observed in Fig. \ref{fig:ene}, large-size coherent vortices emerge far away from the inlet, which interrupt the enstrophy cascade. At these distances, turbulence becomes more intermittent as velocity fluctuations are localized in the vortexes. 

\subsection{Effect of $Re$ and $Cu$ in $2.5$D turbulence}
\label{sec:resReCu}
\begin{figure}
    \centering
    \includegraphics[scale=0.27,keepaspectratio]{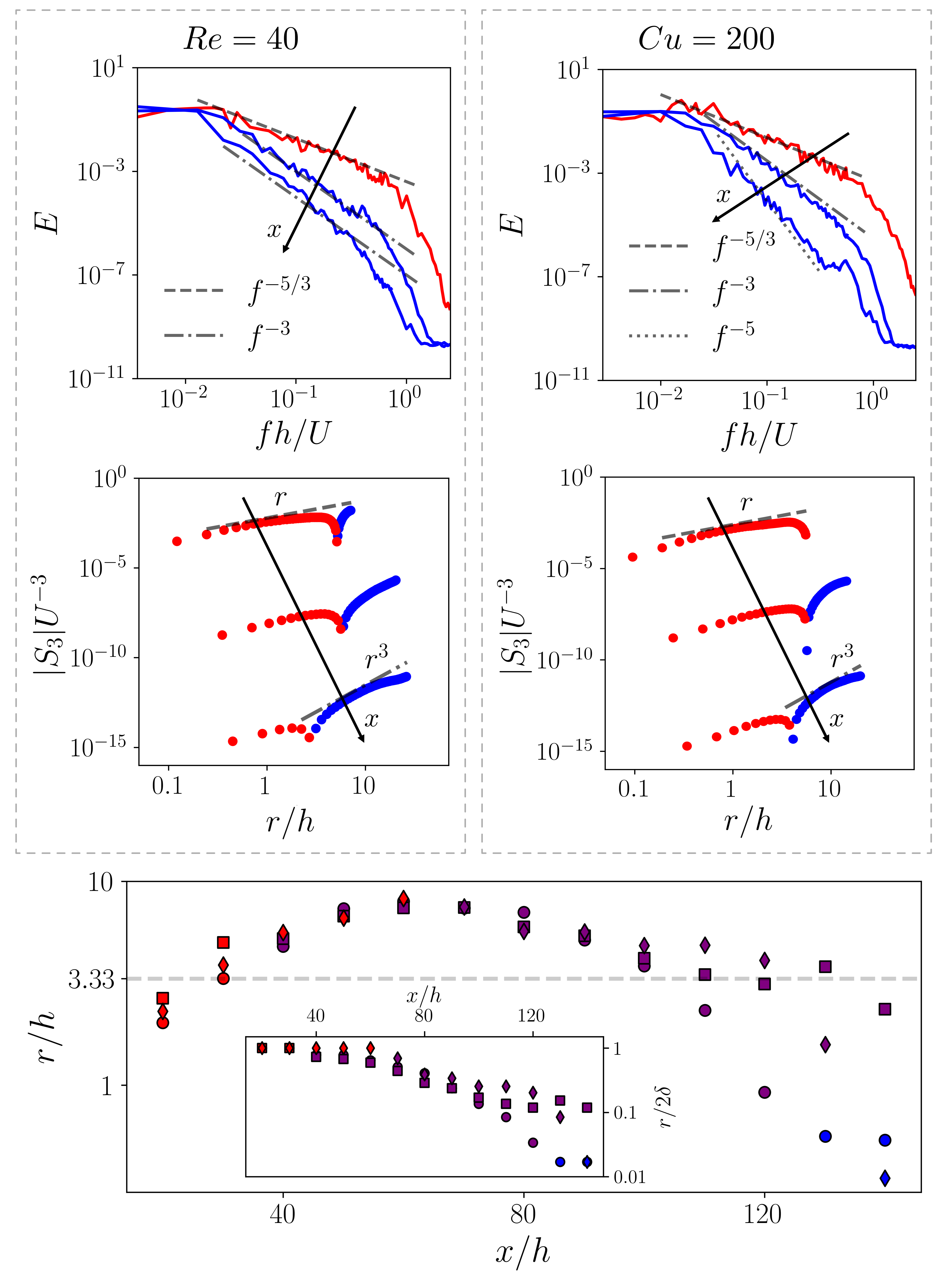}
    \put(-330,440){\large \textit{a})}
    \put(-160,440){\large \textit{b})}
    \put(-330,290){\large \textit{c})}
    \put(-160,290){\large \textit{d})}
    \put(-330,143){\large \textit{e})}
    \caption{Effect of $Re$ and $Cu$ in the mixed-dimensional turbulent planar jet ($L_z = 3.33h$). Panels \textit{a} and \textit{b} report the power spectrum computed at the jet centerline at distances $x = 40h, 80h, 120h$ from the inlet. The 3D turbulence scaling $f^{-5/3}$ (red) and the 2D turbulence scaling $f^{-3}$ (blue) are reported for reference. Panels \textit{c} and \textit{d}: third-order structure function $S_3$ at increasing stream-wise distances, $x = 40h, 80h, 120h$. Red markers identify three-dimensional turbulence and blue markers two-dimensional turbulence. The expected scalings, $r$ (3D) and $r^3$ (2D) are reported for reference. Panel \textit{e}:  transition scale $r$ at which $S_3$ changes sign as a function of the stream-wise position. Markers identify the different cases: squares for $Re=40$, diamonds for $Cu=200$ and circles for the reference case ($Re=20$, $Cu=100$). The color of the marker defines the dimensionality of the flow: red for $3$D ($S_3<0$), blue for $2$D ($S_3>0$) and purple for $2.5$D ($S_3$ changes sign). The inset reports the transition scale $r$ normalized by the local jet thickness, $2\delta$.}
    \label{fig:ene-str_recu}
\end{figure}
In the previous section we observed that constraining the flow in the vertical direction leads to a transition in the turbulent flow, from three-dimensional turbulence (large $L_z$), to a mixed $2.5$D state (intermediate $L_z$) and lastly to two-dimensional turbulence (low $L_z$). We now consider an intermediate case, $L_z=3.33h$, characterized by mixed-dimensional (2.5D) turbulence and investigate the sensitivity of the flow dimensionality on the problem parameters. 
We separately test the effect of a flow parameter, the inlet Reynolds number $Re$, and of the fluid rheology. Specifically, we consider a case at a higher Reynolds number, $Re=40$, and a case at a higher Carreau number, $Cu=200$. An increase in either parameter results in a higher local Reynolds number. The increase in the inlet Reynolds number is attained by halving the reference, zero-shear viscosity $\mu_0$ (the infinite-shear viscosity is reduced accordingly, $\mu_\infty=\mu_0/50$), whereas the Carreau number is increased by doubling the fluid consistency index $\lambda$ with all other parameters left unchanged. The effect of the power-law index $n$ was not tested, as a reduction of $n$ leads to negligible changes in the fluid rheology and an increase of $n$ leads to a laminarization of the jet fluid, due to an increase in the local viscosity. An increase in the Carreau number shifts the transition towards the infinite-shear viscosity at lower values of the local shear rate $\dot{\gamma}$.

We report data from these two additional cases in Fig.~\ref{fig:ene-str_recu}, with $Re=40$ in the left column and $Cu=200$ in the right column. 
The turbulent kinetic energy power spectrum, panels \textit{a} and \textit{b}, shows similar results to the reference case ($L_z=3.33h$, $Re=20$ and $Cu=100$): three- and two-dimensional regimes coexist (large 2D scales and small 3D scales). Near the inlet, $x=40h$, we observe three-dimensional turbulence characterized by the $f^{-5/3}$ scaling of the energy spectrum; as we move downstream, the vertical constraint forces turbulence to the two dimensional regime, as indicated by the scaling $f^{-3}$. 
\revb{For the case at larger Carreau number, $Cu=200$, we however observe a steepening of the power spectrum away from the inlet: at $x=120h$, the power spectrum approaches the scaling $f^{-5}$, similarly to what observed in the more constrained cases. }
The cases at either higher $Re$ or $Cu$ are characterized by a widening of the region where the power-law scalings are observed: the wider separation of scales is owed to the higher value of the local Reynolds number. 
Also the third-order structure functions $S_3$, reported in panels \textit{c} and \textit{d} at three different stream-wise positions, show that the flow is characterised by a mixed-dimensional regime, similarly to what observed for the reference case, Fig.~\ref{fig:strfun32}\textit{a}. We observe that for the $Cu=200$ case the onset of the mixed-dimensional regime occurs at a larger distance from the inlet: at $x=40h$ the structure function is negative at each scale, thus denoting the absence of $2$D turbulence. At further downstream positions, both the $Re=40$ and $Cu=200$ cases show a transition between $2$D and $3$D very similar to the reference case.

The scale at which we observe transition between two- and three-dimensional turbulence is reported in Fig.~\ref{fig:ene-str_recu}\textit{e} for different stream-wise positions. This transition scale is defined as the scale $r$ where the third-order structure function changes sign, from $S_3<0$ (3D) to $S_3>0$ (2D), and corresponds to the largest $3$D scale attained at the specific stream-wise position. We use markers to differentiate the various cases (squares for $Re=40$, diamonds for $Cu=200$ and circles for the reference case) and colors to identify the regime: red for $3$D turbulence ($r$ corresponds to the largest scale, $2\delta$), purple for $2.5$D ($r$ corresponds to the transition scale) and blue for $2$D ($r$ corresponds to the smallest scale considered).
As seen in panel \textit{d}, the case at higher $Cu$ shows a later transition, at $x > 60h$, whereas the case at higher $Re$ exhibits very similar transition length-scale to the reference case. The transition from 3D to 2.5D is initiated at a scale similar to twice the jet thickness, $r\approx 1.5\delta$, as shown in the inset in Fig.~\ref{fig:ene-str_recu}\textit{e}, and as we move downstream it becomes much smaller than the jet thickness. For the reference and $Re=40$ cases we observe a non-monotonic trend in the transition scale $r$ (purple markers), initially increasing up to $x=60h$ and decreasing past this stream-wise position. The decreasing pattern for $x\geq 70h$ is shown as well by the $Cu=200$ case, which instead shows only the monotonically decreasing trend (for $x\geq 70h$, the first stream-wise position at which mixed-dimensional turbulence is reported). The maximum transition scale, achieved at about $x=70h$, is about $3L_z$ for all cases.
We report as well a difference in the transition from $2.5$D to $2$D among the three cases: the flow becomes two-dimensional for both the $Cu=200$ (at $x=140h$) and the reference case (at $x=130h$). The $Re=40$ remains instead characterized by mixed-dimensional turbulence and does not show a transition to $2$D turbulence within the length of the simulation domain. Nevertheless, we observe that  the size of the transition scale $r$ decays with the stream-wise position, thus indicating that eventually the flow will become two-dimensional at a large enough distance from the inlet. We attribute the difference in the $2.5$D to $2$D transition to the lower viscosity of the fluid: the $Re=40$ is characterized by a lower reference zero-shear (and as well, infinite-shear) viscosity, thus delaying the transition to $2$D turbulence.

\section{Conclusions}
We have studied via direct numerical simulations how the vertical confinement of a turbulent planar jet can alter its dimensionality. We show that under the right constraint, a mixed-dimensional turbulent regime appears, that is characterized by the simultaneous presence of large-size two-dimensional and small-size three-dimensional scales. The onset of this particular regime is dictated by the size of the constrained dimension $L_z$: as soon as the flow scales become larger than $L_z$, two-dimensional flow structures appear. This transition is postponed further downstream as the flow is less constrained (increasing $L_z$). Therefore, for sufficiently large $L_z$, turbulent scales are simply three-dimensional along the jet, and the direct cascade of energy is enabled. Conversely, a strong confinement (small $L_z$) makes the flow two-dimensional: the direct cascade of energy is disrupted and the direct cascade of enstrophy takes place. Both cascades are conserved wherever the mixed-dimensional turbulent state is present: the direct cascade of energy is active at small scales, whereas the direct cascade of enstrophy dominates at large scales. 
The generality of these findings has been tested: two additional cases, one at a higher inlet Reynolds number and one at a higher Carreau number, have been investigated for the mixed-dimensional case, $L_z=3.33$. We report that the mixed-dimensional 2.5D turbulent regime is still observed at both higher $Re$ and higher $Cu$: the flow is still characterized by large two-dimensional and small three-dimensional scales. The largest transition scale, marking the largest 3D turbulent scales, is attained at $x=70h$ for the three cases ($Re=40$, $Cu=200$ and reference case). Minor differences are observed in the transition from 3D to 2.5D, with the $Cu=200$ case showing 2.5D turbulence further downstream, and in the transition from 2.5D to 2D, which does not occur within the computational domain for the $Re=40$ case.
\reva{We expect that considering a turbulent Newtonian jet (at a comparable inlet local Reynolds number) would have similar effects of increasing the inlet Reynolds number (case at $Re=40$, see figure~\ref{fig:ene-str_recu}\textit{e}): the region characterized by $2.5$D turbulence widens and the transition to $2$D flow shifts downstream. A Newtonian fluid lacks any shear-thinning properties, thus it is not characterized by a local viscosity increasing with the stream-wise position, thus allowing smaller turbulent structures to exist over a longer distance from the inlet. To better quantify this effect, a complete comparison with Newtonian turbulence is required, and may be the object of a future study.
In addition, when increasing either the Reynolds or the Carreau number,} we found that the mixed-dimensional configuration is an overall more energetic state, thus partially retaining the three-dimensionality in the flow while deferring the emergence of two-dimensional strong vortical structures further downstream from the inlet. 
The direct enstrophy and energy cascades and the respective scalings we report here are the same of a Newtonian fluid: the Carreau fluid is a non-Newtonian fluid model characterized by shear-thinning alone, which allows to attain Newtonian turbulence at a relatively low Reynolds number. 

\backsection[Acknowledgements]{The research was supported by the Okinawa Institute of Science and Technology Graduate University (OIST) with subsidy funding from the Cabinet Office, Government of Japan. M.E.R. also acknowledges funding from the Japan Society for the Promotion of Science (JSPS), grant 24K17210. The authors acknowledge the computer time provided by the Scientific Computing and Data Analysis section of the Core Facilities at OIST, and the computational resources on SQUID provided by the Cybermedia Center at Osaka University through the HPCI System Research Project (project ID: hp230018).}

\backsection[Declaration of interests]{The authors report no conflict of interest.}

\backsection[Author ORCIDs]{\\
Christian Amor, \url{https://orcid.org/0000-0002-9710-7917} \\
Giovanni Soligo, \url{https://orcid.org/0000-0002-0203-6934} \\
Andrea Mazzino, \url{https://orcid.org/0000-0003-0170-2891} \\
Marco Edoardo Rosti, \url{https://orcid.org/0000-0002-9004-2292}}

\appendix
\section{Structure function in planar jets}
\label{app:strucfun}

The $p$-th order structure function is defined as the $p$-th moment of velocity differences \citep{frisch1995turbulence}:
\begin{equation}
   S_p(r) = \langle \left( \Delta u \right)^p \rangle.
   \label{eq:s3app}
\end{equation}
Angle brackets indicate averaging in time, in the vertical direction $z$ and over velocity differences separated by the same distance $r=\left|\boldsymbol{r}\right|$.

\begin{figure}
   \centering    
   \includegraphics[scale=0.25,keepaspectratio]{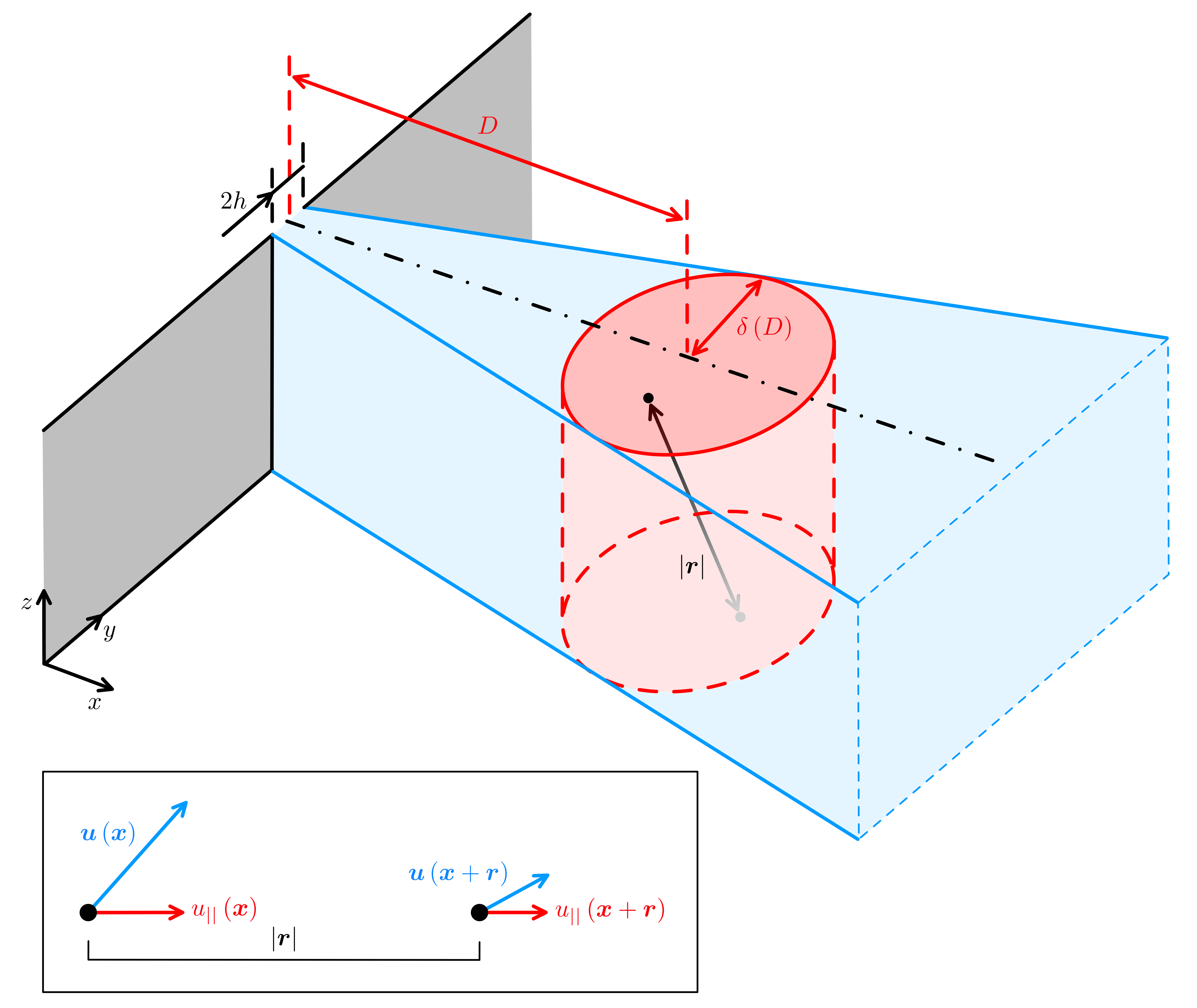}
   \caption{Calculation of the differences of velocity $\Delta u$ in the planar jet. The differences of longitudinal velocity $u_{||}$ between points separated by a distance $r$ is computed among points within the red cylinder.}
   \label{fig:sketch}
\end{figure}

The structure function is computed within a cylinder (in red in the sketch in Fig. \ref{fig:sketch}) at a set distance $D$ from the inlet. The cylinder axis is aligned with the vertical direction $z$ and its radius is equal to the jet thickness at distance $D$ from the inlet, i.e. $\delta(D)$. The jet thickness $\delta$ is defined as the distance from the centerline at which the streamwise velocity equals half of the centerline velocity. Longitudinal velocity differences,
\begin{equation}
   \Delta u(r) =  u_{||} \left( \boldsymbol{x} + \boldsymbol{r} \right) - u_{||} \left( \boldsymbol{x} \right) = \left( \boldsymbol{u} \left( \boldsymbol{x} + \boldsymbol{r} \right) - \boldsymbol{u} \left( \boldsymbol{x} \right) \right) \cdot \boldsymbol{r} / \left| \boldsymbol{r} \right|,
   \label{eq:veldiff}
\end{equation}
are computed across points located at distance $r$ lying within the cylinder; structure function data are then averaged over time and over \revb{couple of points separated by the same} separation distance $r$ (angle brackets in Eq.~\ref{eq:s3app}). 
\reva{As the slenderness of the cylinder changes among different vertical constraints and different stream-wise positions (for instance, slender cylinder for $x=40h$ and $L_z=13.33h$, or stocky cylinder for $x=120h$ and $L_z=0.83h$), we investigated the effect of different cylinder aspect ratios on the computed structure function. The third-order structure function was thus recomputed considering only couple of points lying on the same $x-y$ plane, thus eliminating the effect of the separation in the $z$ direction. Direct comparison among the structure function computed within a cylinder and computed on $x-y$ planes showed minimal differences in the values of the third-order structure function and no difference in its sign.}

The jet thickness $\delta$ increases with the distance from the inlet $D$: as we move away from the inlet, the maximum separation distance, $2\delta$, increases as well. The maximum separation distance is divided in $N_r=60$ uniformly spaced bins; the width of the bins is thus proportional to the jet thickness, and generally increases with the distance from the inlet $D$. The smallest width of the bin, found close to the inlet, is about the same size of the grid spacing; as one moves away from the inlet the width of the bin increases, resulting in a limited loss of spatial resolution at the smallest scales (nonetheless smaller than ten grid spacings at the farthest distance considered in the present work). Therefore, in some cases very small scales may still show three-dimensional turbulence (and a direct energy cascade), however they may not be fully detected by the structure function, as they are averaged over the smallest separation distance considered.

In this study, we utilize the third-order structure function $S_3$, which we show to be a reliable tool to inspect the flow dimensionality and the preferred turbulent cascade process scale-by-scale. The first derivation of $S_3$ resulted in the celebrated Kolmogorov's 4/5th law in three-dimensional turbulence \citep{kolmogorov1991dissipation}: $S_3(r) = -\frac{4}{5} \epsilon r$, where $\epsilon$ is the mean energy dissipation per unit mass. The $S_3$ laws in two-dimensional turbulence were derived in the late 90's \citep{lindborg1999can,bernard1999three}: for the inverse energy cascade, $S_3(r) = \frac{3}{2} P r$ (where $P$ is the mean energy injection per unit mass), and for the direct enstrophy cascade, $S_3(r) = \frac{1}{8} \zeta r^3$ (where $\zeta$ is the mean enstrophy dissipation per unit mass). \cite{cerbus2017third} related $S_3(r)$ to the flux functions:
\begin{equation}
   S_3(r) = -\frac{3}{2} \Pi \left(\frac{a}{r} \right) r + \frac{1}{8} Z \left(\frac{a}{r} \right) r^3 + \cdots,
\label{eq:S3}
\end{equation}
with $\Pi$ and $Z$ being the fluxes of energy and enstrophy, respectively, and $a$ an $O(1)$ numerical constant (the quantity $a/r$ can be thought of as a wavenumber, $k\sim a/r$). 
Equation \ref{eq:S3}  expresses $S_3(r)$ as a combination of the energy and enstrophy fluxes and its sign is affected by values of both fluxes. The previous laws for $2$D turbulence can be easily derived from Eq. \ref{eq:S3} \citep{cerbus2017third}.

\bibliographystyle{jfm}
\bibliography{totalbib} 

\end{document}